\documentclass[longbibliography,aps,pre,twocolumn,groupedaddress]{revtex4-1}
\usepackage{graphicx}
\usepackage{bm}
\usepackage{amsmath}
\usepackage{dcolumn}

\begin{document}

\title{Critical Phenomena in Quasi-Two-Dimensional Vibrated Granular Systems}

\author{Marcelo Guzm\'an and Rodrigo Soto}

\affiliation{Physics Department,  Facultad de Ciencias F\'\i sicas y Matem\'aticas, Universidad de Chile, Santiago, Chile.}

\date{\today}

\begin{abstract}
The critical phenomena associated to the liquid to solid transition of quasi-two-dimensional vibrated granular systems is studied using molecular dynamics simulations of the inelastic hard sphere model. The critical properties are associated to the fourfold bond-orientational order parameter $\chi _4$, which measures the level of square crystallization of the system. Previous experimental results have shown that the transition of $\chi_4$, when varying the vibration amplitude, can be either discontinuous or continuous, for two different values of the height of the box. 
Exploring the amplitude-height phase space, a transition line is found, which can be either discontinuous or continuous, merging at a tricritical point and the continuous branch ends in an upper critical point.  In the continuous transition branch, the critical properties are studied. The exponent associated to the amplitude of the order parameter is $\beta=1/2$, for various system sizes, in complete agreement with the experimental results. However, the fluctuations of $\chi _4$ do not show any critical behavior, probably due to crossover effects by the close presence of the tricritical point. 
Finally, in quasi-one-dimensional systems, the transition is only discontinuous, limited by one  critical point, indicating that two is the lower dimension for having a tricritical point.

\end{abstract}

\pacs{}

\maketitle

\section{Introduction} \label{sec.intro}

The study of granular matter have attracted a large attention not only because of its numerous applications to describe natural and industrial processes, but also because it serves as an excellent prototype of non-equilibrium systems, where it is possible to test different hypothesis and models. For example, and to simply mention a few, it has been possible to study in detail the fluctuation theorems~\cite{aumaitre2001power, PhysRevLett.92.164301, PhysRevLett.108.210604}, the extension of kinetic theory when the spatio-temporal scales are not completely separated~\cite{kadanoff1999built,goldhirsch2003rapid,PoschelKT}, the effect of correlations in the development of giant density fluctuations~\cite{Narayan105}, or the formation of patterns and structures~\cite{AransonPattern}.

Of particular interest is the quasi-two-dimensional (Q2D) geometry, where grains are placed in a shallow box, which is  vertically vibrated. Here, if the box height is smaller than two grain diameters, it is possible to follow experimentally the motion of all grains. Together with the possibility to manipulate the interparticle interactions, this access to the global response and motion at the grain scale, make this geometry particularly relevant to build the statistical thermodynamics of non-equilibrium systems~\cite{Review2016}. Q2D systems have been extensively analyzed both numerically and through simulations~\cite{Olafsen,Losert,Prevost,Melby,Reyes,Clerc,Argentina}. In the pioneer works of Olafsen and Urbach~\cite{Olafsen}, and Losert \textit{et al.}~\cite{Losert}, it was shown that a Q2D granular gas presents both clustering and ordering transitions for low vibration amplitudes. For large vibration amplitudes, it was shown that when the vibration amplitude or filling density surpasses  a certain threshold, a solid-liquid-like transition takes place, and furthermore, different solid phases appear depending  on the filling fraction and box height~\cite{Prevost,Melby}.
This phase separation is produced by a negative compressibility in the associated 2D state equation and it was shown that in the transition, the pressure as a function of the density reaches a plateau as in the van der Waals case~\cite{Clerc}. 
Other aspects of the dynamics of Q2D as  the effect of forcing, dissipation, and inelasticity, together with the implementation of effective 2D models have been studied in detail as well (see review~\cite{Review2016} and references therein).

Using two Q2D configurations of different heights and global densities, it was shown experimentally that the liquid to solid transition could be either continuous or discontinuous for the crystalline order parameter when increasing the vibration amplitude~\cite{PhysRevLett.109.095701}. In the continuous case, five critical exponents were measured, which present universality properties when compared to other experiments where the plate mechanical properties were changed~\cite{PhysRevLett.109.095701,PhysRevE.91.012141}.
Our objective in this article is twofold: on one hand, we aim to reconcile the fact that the transition has two different characters  when changing the height and, on the other hand, to test the universality of the exponents found experimentally. To do so, we analyze the system through molecular dynamics (MD) simulations~\cite{MRC,poschel2005computational}. This approach has an inherent advantage: the parameters,  particularly the height of the box, can be varied continuously unlike in the experimental counterpart. It is found that a tricritical point appears in the amplitude--height parameter space, where the continuous and discontinuous transitions converge.
The universality is analyzed by considering dissipation coefficients that are quite different to those used experimentally.
We observe, also, that two is the lower critical dimension for the existence of the tricritical point as quasi-one-dimensional systems do not show continuous transitions.

The plan of the paper is as follows. In Section~\ref{sec.setup} we describe the configuration under study, the order parameter that characterizes the liquid to solid transition and its main properties. Section~\ref{sec.simuq2d} describes the simulation method and parameters, and presents the results for quasi-two-dimensional systems. The case of  quasi-one-dimensional systems, where larger wavelengths can be achieved, is analyzed in Section~\ref{sec.simuq1d}. Finally, a discussion on the results is given in Section~\ref{sec.discussion}.

\section{Liquid to solid transition in Q2D systems} \label{sec.setup}

\begin{figure}[htb]
 \includegraphics[width=\columnwidth]{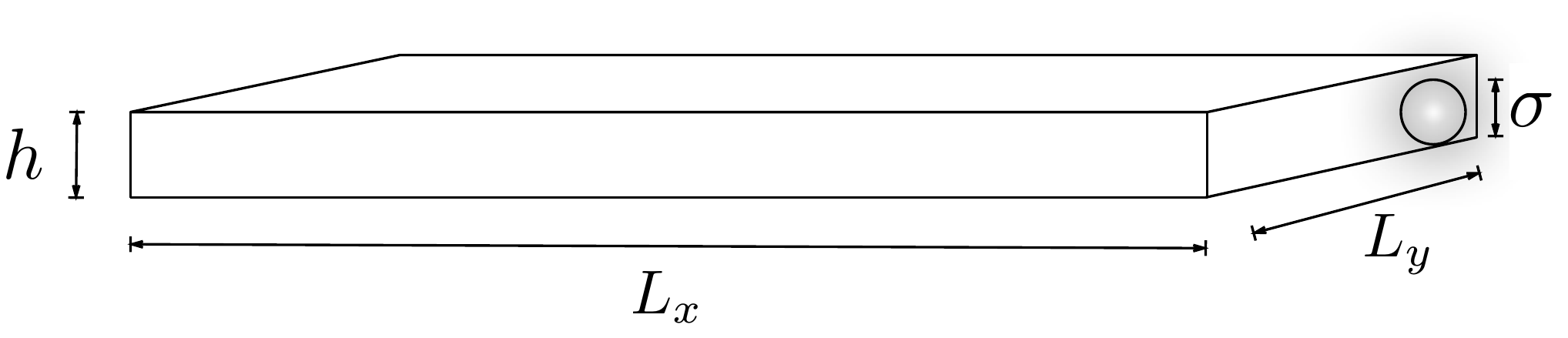}
 \caption{Shallow box system of lateral dimensions $L_x,L_y\gg \sigma$ and height in the range $\sigma<h<2\sigma$. A grain is shown as reference. The whole box is vibrated vertically with amplitude $A$ and angular frequency $\omega$ in presence of gravity.\label{Q2D system}}
 \end{figure}

Figure~\ref{Q2D system} presents the quasi-two-dimensional geometry under study. $N$ monodisperse spherical grains of diameter $\sigma$ are placed in a shallow box of large lateral dimensions, $L_x,L_y\gg \sigma$, while the height is limited to the range $\sigma<h<2\sigma$. The whole box is vibrated vertically with angular frequency $\omega$ and amplitude $A$, in presence of a gravity acceleration $g$. In experiments, the oscillation waveform is sinusoidal, while in simulations  a bi-parabolic waveform is used for higher accuracy \cite{godoy2008rise}. The collisions between grains and with the top and bottom walls are inelastic and frictional. For fixed geometrical and mechanical parameters, and keeping constant the frequency, a transition takes place when increasing the amplitude. Below the threshold amplitude a homogeneous (except for boundary effects near the lateral walls) fluid-like state develops and above this threshold, a solid-like cluster forms surrounded by the liquid phase. Depending on the height and the amplitude of oscillation, the solid cluster presents  crystalline phases of different symmetries~\cite{Melby}. In the range of parameters used in Refs.~\cite{PhysRevLett.109.095701,PhysRevE.91.012141}, the crystal consists on two intercalated layers of square symmetry.  

Experimentally, the density fluctuations did not reveal any critical behavior near the transition. However, the transition manifests when analyzing the fourfold bond-orientational order parameter: for each particle $j$, we compute
\begin{equation}
\chi_4^j=\frac{1}{N_j}\sum_{s=1}^{N_j}e^{4i \alpha _s^j},
\end{equation}
where $N_j$ is the number of nearest neighbors and $\alpha_s^j$ is the angle of the two-dimensional projection of the relative vector $\textbf{r}_s-\textbf{r}_j$ with respect to an arbitrary fixed axis. Note that $0\le|\chi_4^j|\le1$, reaching its maximum value when the particle is in a perfect square lattice. The time average of the module of $\chi _4$ 
\begin{equation}
\left<|\chi _4| \right>=\left< \frac{1}{N}\sum_{j=1}^N|\chi_4^j|\right>,
\end{equation}
computed in the steady state, is an order parameter that measures the fraction of particles in the ordered phase. Two configurations were used in Ref.~\cite{PhysRevLett.109.095701}:  C1, with $h=1.83\sigma$, and C2, with  $h=1.94\sigma$. In both cases, for amplitudes larger than a threshold, $\left<|\chi _4 |\right>$ increases its value. For C1 there is a discontinuous jump, while for  C2 the order parameter changes continuously although with discontinuous (apparently diverging) derivative. 

Below the threshold amplitude, still in the liquid phase, small crystalline patches with square symmetry, of finite size and lifetime, coexist  with the liquid environment. Their existence is evidenced by the analysis of the Fourier components of $\chi _4$, \begin{equation}
\widehat{\chi}_4\left(\bm k,t \right)=\sum_{j=1}^N\chi_4^je^{i\bm k \cdot\bm r_j(t)},
\end{equation}
where their fluctuations are computed with the fourfold bond-orientational structure factor
\begin{equation}
S_4(\bm k)=\frac{\left<|\widehat{\chi}_4(\bm k,t)-\left< \widehat{\chi}_4(\bm k,t)\right>|^2 \right>}{N}.
\end{equation}
For both configurations, it was found that $S_4$ showed an Ornstein-Zernike-like behavior in the limit $k\sigma\ll1$, $S_4(k)\approx S_4(0)/[1+(\xi _4 k)^2]$, where  $\xi_4$ is the fourfold correlation length and $S_4(0)$ is the associated static susceptibility. While no critical behavior was found for C1, for C2 two critical exponents were found, associated to the divergence of $\xi_4$ and $S_4(0)$ at the transition.

\section{Simulations of quasi-two-dimensional systems} \label{sec.simuq2d}

We study the system through three-dimensional MD simulations, using the inelastic hard sphere model \cite{MRC,poschel2005computational}, with identical spherical grains and using periodic boundary conditions for the lateral walls. 
The fixed parameters of the simulation are the Q2D number density $\phi _{2D}\equiv N\sigma^2/L_xL_y=0.9875$, with $L_x=L_y$, and the normalized frequency of oscillations of the container $\omega \sqrt{\sigma/g}=5$. Also fixed are the grain-grain and grain-wall friction coefficients $\mu = 0.03$ and restitution coefficients $\alpha=0.998$, respectively. These values were chosen by inspection to ensure the appearance of clusters with square symmetry in the range of heights $1.73\sigma\leq h\leq 1.85\sigma$. 

We remark that the friction coefficients chosen in this work are one order of magnitude below the experimental values. This difference has its origin in that in simulations, particles are perfectly spherical and the plates are also perfectly flat, contrary to experiments, where slight roughness and imperfections are present. Hence, in simulations using dissipation coefficients similar to the experimental ones, the particles reach states with no horizontal motion~\cite{AbsorbingState}.
For a quantitative comparison with experiments at comparable densities, simulations had to include explicitly these effects, which allow to achieve fluidized states in experiments even for low particle concentrations~\cite{PhysRevLett.106.088001}.  

\begin{figure}[htb]
 \includegraphics[width=0.7\columnwidth]{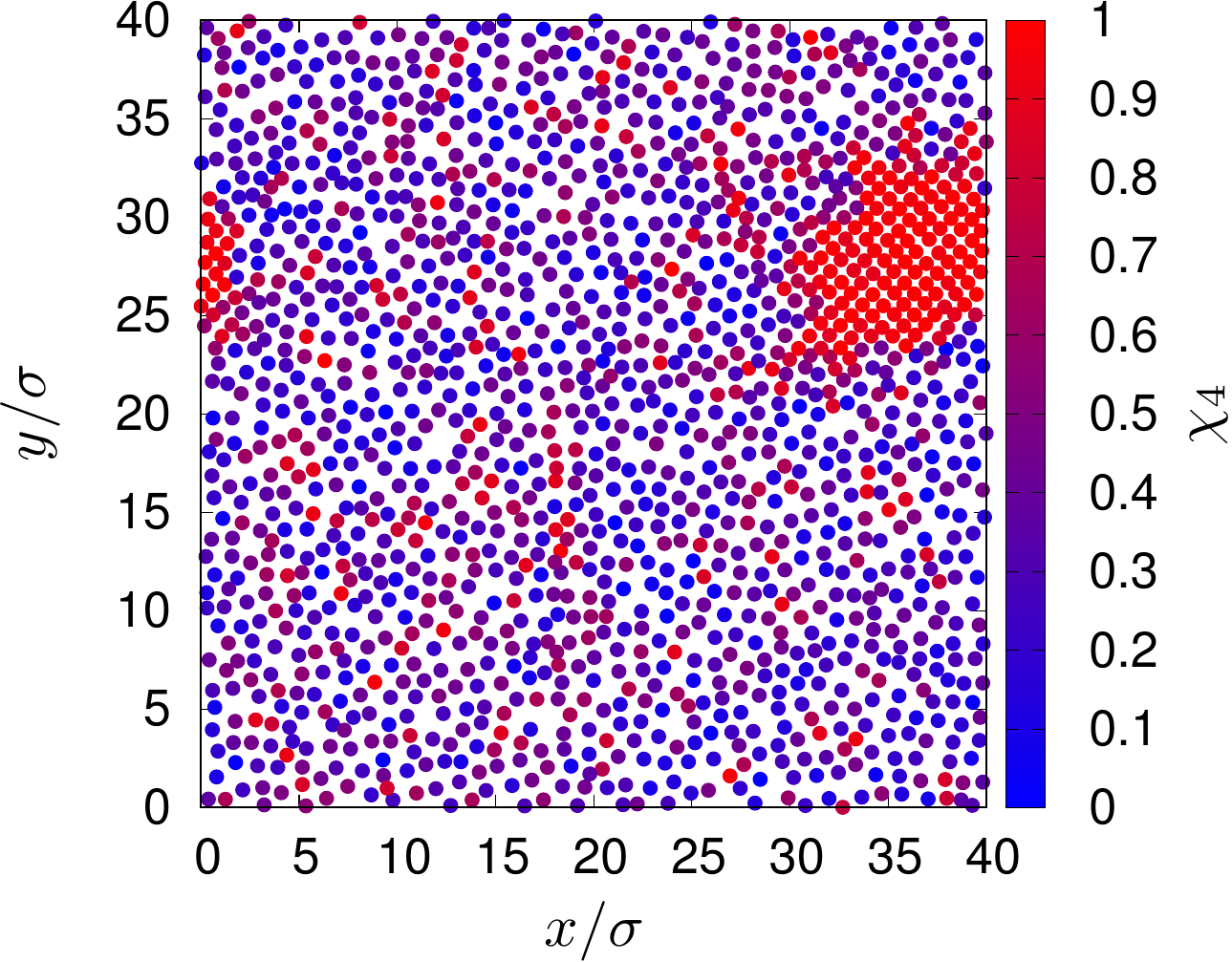}
 \caption{Cluster with square symmetry obtained in simulations in a system of $N=1580$ particles in box with lateral size $40\sigma\times40\sigma$ and height $h=1.8\sigma$. The amplitude is $A=0.2\sigma$. The color code indicates the absolute value of $\chi_4$ for each particle. Grains have been drawn at a smaller size, with diameter $\approx 0.8\sigma$, to appreciate the crystalline structure of the cluster. Had they been depicted with their real size, the two layers would have overlapped when projected in 2D~\cite{Prevost}.}
\label{fig.squarecluster}
 \end{figure}

Nonetheless, using perfect spheres and flat walls, our simulations reproduce the geometrical properties of the solid cluster and are therefore appropriate for the purposes declared in the Introduction. Figure \ref{fig.squarecluster} shows a cluster in its stationary regime. Its size and shape remain approximately constant unlike its orientation, which displays Brownian rotation.

\subsection{Fourfold Bond-Orientational Parameter: Phase Space}

\begin{figure}[htb]
 \includegraphics[width=0.8\columnwidth]{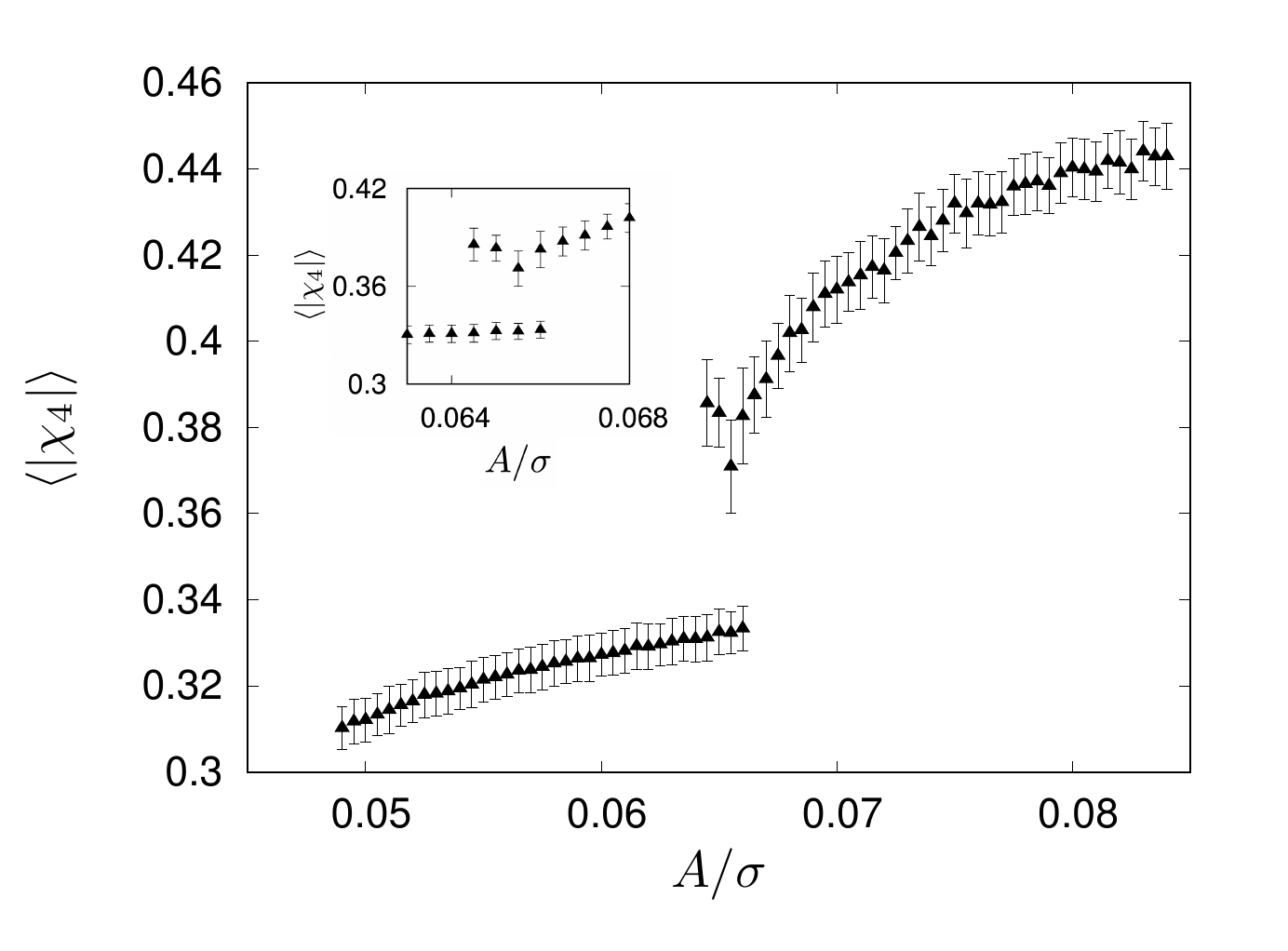}\\
 \includegraphics[width=0.73\columnwidth]{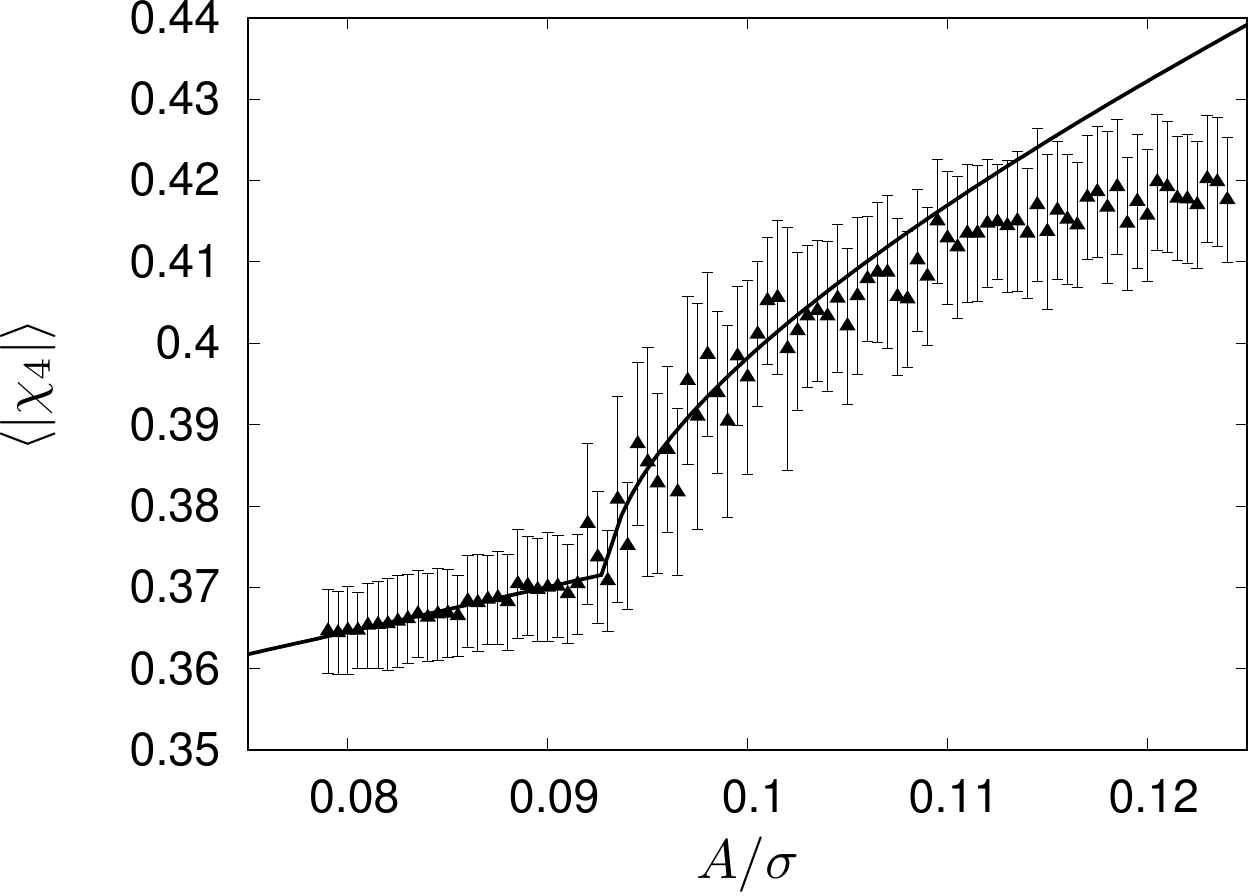}
 \caption{Liquid to solid transition as evidenced by the order parameter $\left<|\chi _4| \right>$ when increasing the amplitude. Above the transition amplitude, stable solid clusters form. The error bars indicate the standard deviation. Top: discontinuous transition for $h=1.74\sigma$, where the inset evidences the existence of bistability. Bottom: continuous transitions for $h=1.8\sigma$. The solid line is the fit close to the transition to determine the critical exponent.} 
 \label{fig.transition}
 \end{figure}

As in the experiments, depending on the height $h$ we found two kinds of transitions for $\left<|\chi _4|\right>$ as a function of the amplitude $A$ (see Fig.~\ref{fig.transition}). For the continuous transition, $\left<|\chi _4|\right>$ can be modeled as $\chi_4^L+\Delta \chi_4$, where $\chi_4^L=aA+b$ is the linear trend observed prior the transition, and 
\begin{align}
\Delta \chi_4=c\left(A-A_c\right)^{\beta} \label{eq.powerlaw}
\end{align} 
is the power-like behavior observed after the transition. Fitting the results to the model as described in Ref.~\cite{PhysRevLett.109.095701}, we obtain $\beta=0.56\pm0.18$ and the non-universal parameters  $a$, $b$, $c$, and $A_c$; the fitted parameters for different heights and system sizes are presented in Table~\ref{table.fitvalues}.

\begin{table}[htb]
\caption{Critical exponent $\beta$ and amplitude $A_c$ for different values of system sizes and box heights in the region of the continuous transition. The values are obtained using the fitting protocol described in Ref.~\cite{PhysRevLett.109.095701}.}
\begin{tabular}{cccD{.}{.}{1.3}D{.}{.}{1.6}D{.}{.}{1.7}}  
$L_x/\sigma$ & $L_y/\sigma$ & $N$ & \multicolumn{1}{c}{$h/\sigma$} & \multicolumn{1}{c}{$\beta$} & \multicolumn{1}{c}{$A_c/\sigma$}  \\
\hline
  40  & 40 & 1580 & 1.8 & 0.56(18) & 0.093(1)\\
  60  & 60 & 3555 & 1.82 & 0.51(1) & 0.094(1)\\
  71  & 71 & 5000 & 1.82 & 0.40(1) & 0.091(1)\\
  71  & 71 & 5000 & 1.825 & 0.44(1) & 0.094(1)\\
  80  & 80 & 6320 & 1.83 & 0.50(1) & 0.095(1)\\
  90  & 90 & 7999 & 1.83 & 0.44(1) & 0.094(1)\\
  90  & 90 & 7999 & 1.84 & 0.52(1) & 0.101(1)\\
 100 & 100 & 9875 & 1.84 & 0.52(1) & 0.099(1) \\ 
 100 & 100 & 9875 & 1.85 & 0.54(2) & 0.109(1) 
\end{tabular}
\label{table.fitvalues}
\end{table}

\begin{figure}[htb]
  \includegraphics[width=\columnwidth]{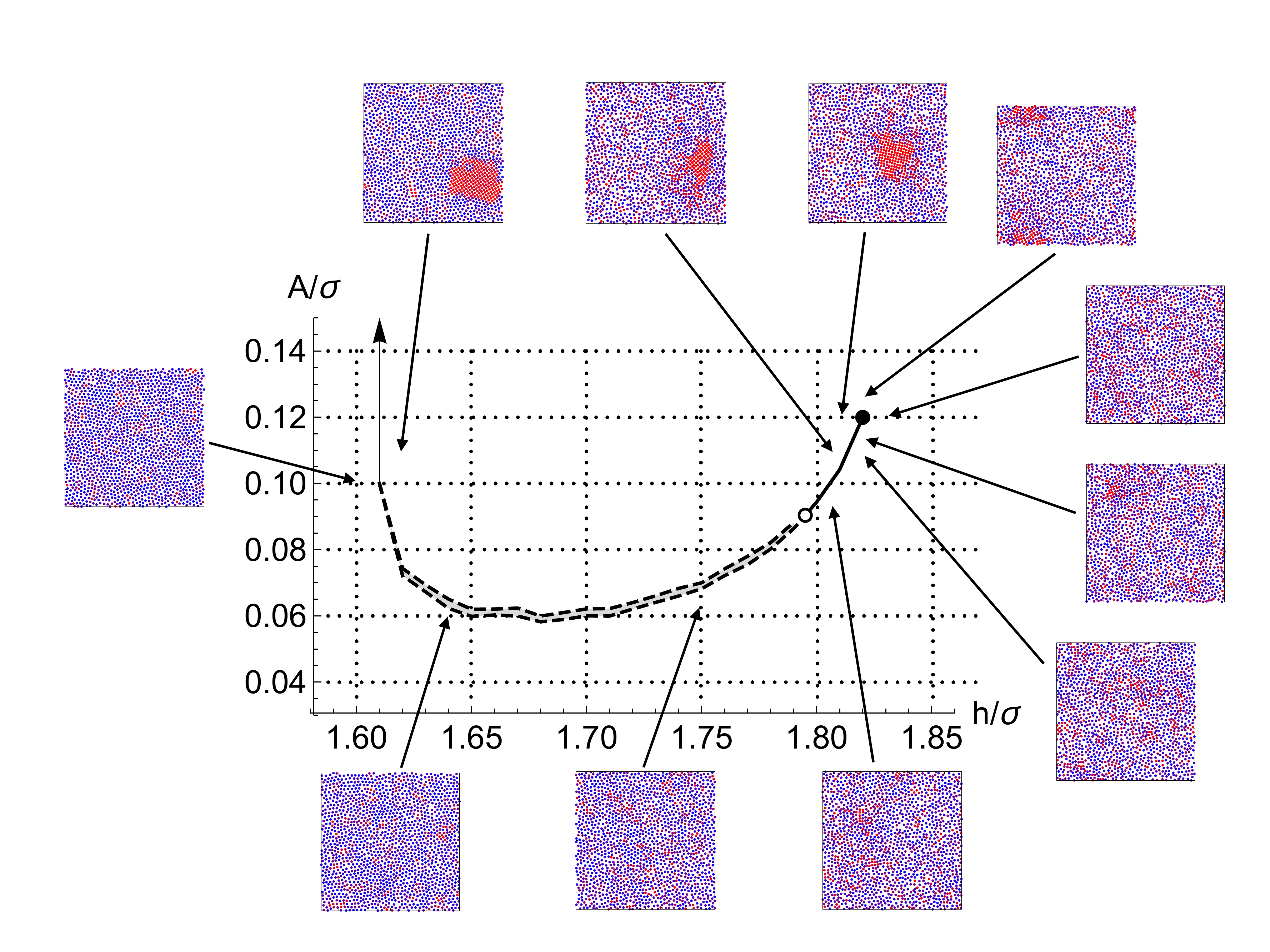}
 \caption{Amplitude--height phase space of the transition for $N=1580$ particles, where the shaded region represents the bistablilty of the system. The dashed lines denote the discontinuous transitions, whereas the solid lines the  continuous ones. The tricritical point is indicated by an empty circle while the upper critical point at the end of the continuous transition by a black circle. The arrow indicates that up to the highest values of $A$ the discontinuous transition is present, without any evidence of a lower critical point. In all cases we explore the phase space until no transition was found.  The position of the tricritical and critical points for other values of $N$ are indicated in Table~\ref{table.criticalpoints}. Typical configurations for special points in the parameter space are displayed.
 }
\label{fig.phasespace}
 \end{figure}

\begin{figure}
 \includegraphics[width=0.7\columnwidth]{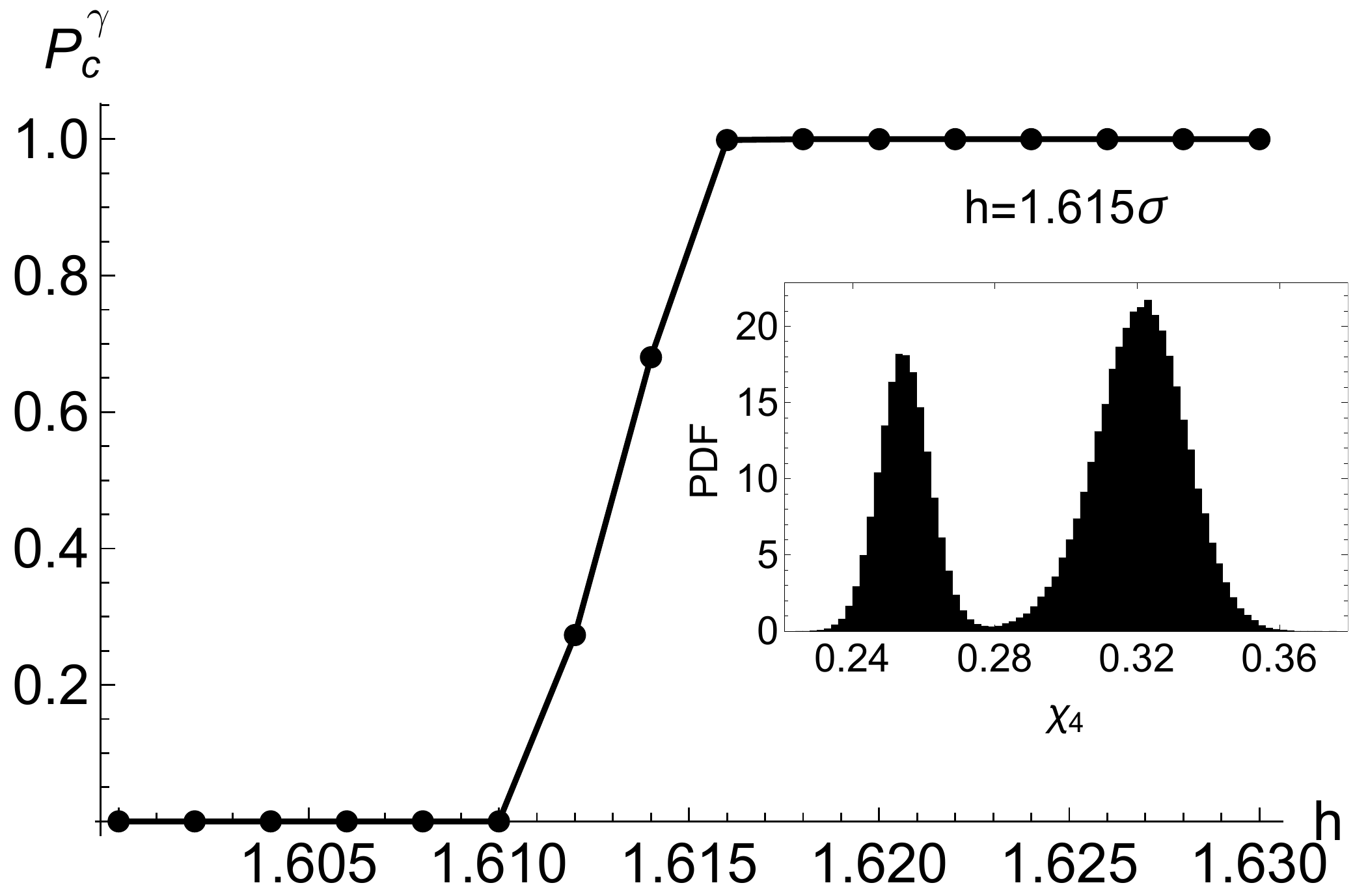}\\
 \includegraphics[width=0.7\columnwidth]{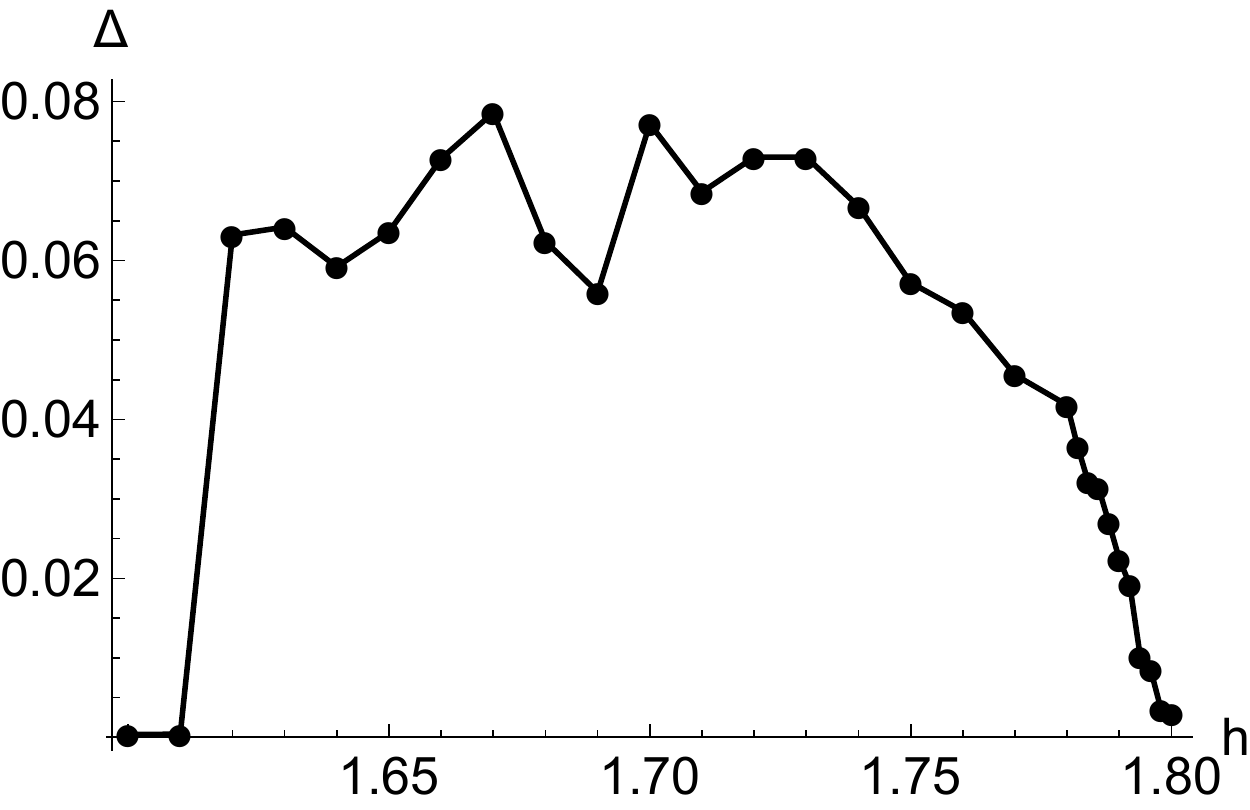}\\
 \includegraphics[width=0.7\columnwidth]{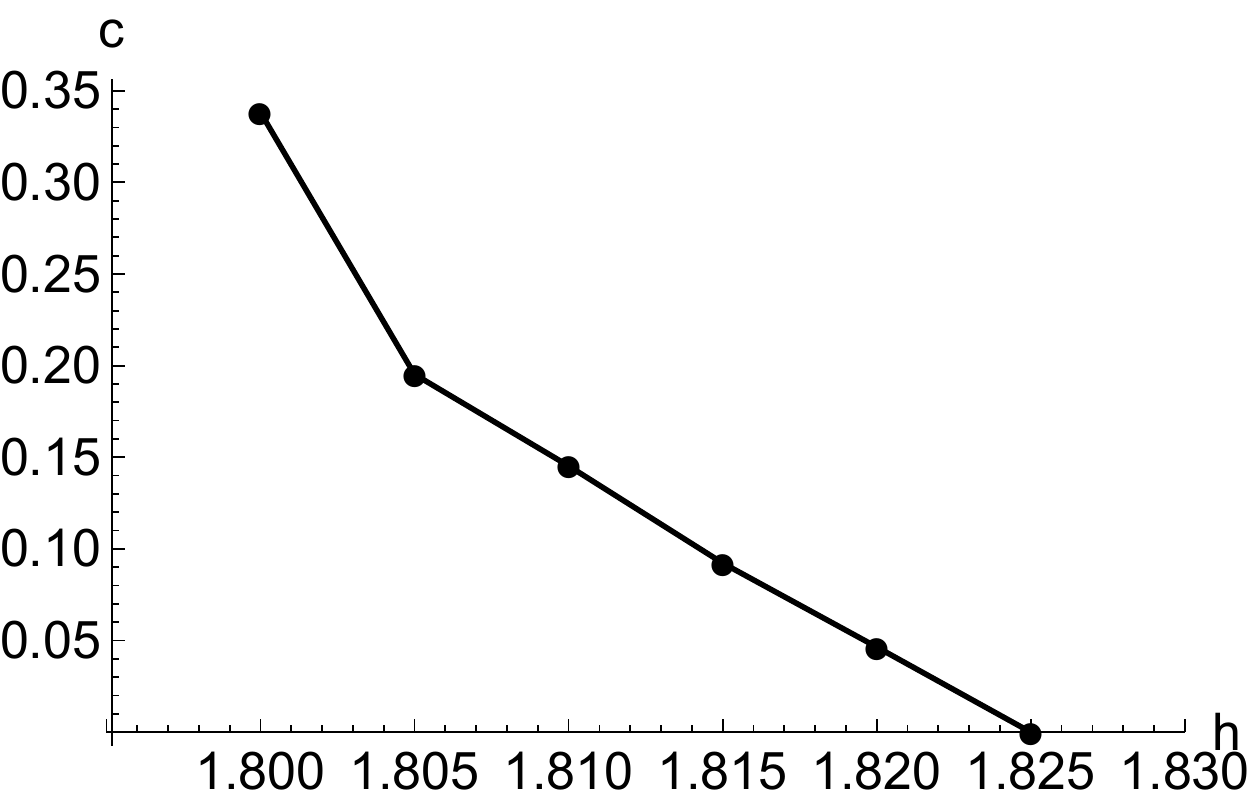}
 \caption{Order parameters $P_c$ to the power $\gamma$ (top), $\Delta$ (middle), and $c$ (bottom) as a function of the box height $h$, for $N=1580$ particles, where $P_c$ has been averaged over the range $A=0.108$ to $A=0.148$. The discontinuous transition line, the tricritical point, and the upper critical point are identified by the vanishing of  $P_c$, $\Delta$, and $c$, respectively. The inset (top) shows the probability density function for $h=1.615$ as an example. The exponent $\gamma$ has been fitted to 1.5, for which $P_c^\gamma$ vanishes linearly. }
 \label{fig.orderparameters}
\end{figure}

Analyzing $\left<|\chi _4|\right>$ it is possible to build the transition diagram in the amplitude--height phase space, which is shown in Fig.~\ref{fig.phasespace} for $N=1580$, together with typical configurations in the vicinity of the transition line. Increasing the amplitude, the liquid to solid transition takes place, where a solid cluster forms. For small heights the transition is discontinuous with a small region of bistability, while for larger heights the transition is continuous. A tricritical point  separates the two cases. In the explored region of parameters, the discontinuous transition does not show a lower critical point and apparently the transition exists up to large values of $A$. The continuous transition, on the other hand, ends in an upper critical point. The position of the transition line and the critical points are identified with adequate order parameters. The discontinuous transitions line is characterized noting that the solid-like cluster remains with finite size until the transition. Computing the probability distribution function for $\left<|\chi _4|\right>$ it is found that it presents two peaks, one corresponding to the homogeneous liquid phase and another, at higher values, associated to the cluster. The order parameter $P_c$ is the area below this second peak (the probability to get a cluster), which vanishes continuously when decreasing $h$ for fixed amplitude, with a power law $P_c\sim(h-h_1)^{0.66}$,  marking the position of the transition line (see Fig.~\ref{fig.orderparameters}-top). The bistability region is recognized by direct observations of the configurations at different instants of time for fixed parameters. 
The position of the tricritical point is determined by analyzing $\Delta$, the jump of $\left<|\chi _4|\right>$ at the discontinuous transition, when increasing $A$ at fixed heights (see Fig.~\ref{fig.transition}-top). Figure \ref{fig.orderparameters}-middle shows $\Delta$, which vanishes at the tricritical point, with a power law $\Delta\sim(h_2-h)$. Finally, the upper critical point that ends the continuous line is determined by the study of the fitting parameter $c$ in Eq.~(\ref{eq.powerlaw}), which measures the amplitude of the ordered phase, and vanishes at the upper critical point as $c\sim(h_3-h)$ (see Fig.~\ref{fig.orderparameters}-bottom).

We also analyzed different system sizes, keeping all the intensive parameters fixed. Similar qualitative behaviors are found up to largest studied case, $N=9875$, finding the same values for the critical exponent $\beta$ and similar values for the two critical points   (see Tables~\ref{table.fitvalues}  and \ref{table.criticalpoints}). In particular, the tricritical and upper critical points remain always at finite distance along the transition line and converge to finite values in the thermodynamic limit.

\begin{table}[htb]
\caption{Position (amplitude $A$ and height $h$) of the tricritical and upper critical points in Q2D systems for different system sizes, indicated by the number of particles $N$. The thermodynamic limit $N=\infty$ is obtained extrapolating all measured values with  $X(N)=X_\infty - X_1/N$.}
\begin{tabular}{p{4em}p{4em} p{4em} p{4em} p{4em} }  
$N$ &  $A_{\text{tri}}/\sigma$ & $h_{\text{tri}}/\sigma$ & $A_{\text{up.cri}}/\sigma$ & $h_{\text{up.cri}}/\sigma$ \\
\hline
 1580 & 0.077  & 1.770  & 0.090 & 1.800\\ 
 3555 & 0.086  & 1.805  & 0.102 & 1.830\\ 
 5000 & 0.088  & 1.815  & 0.105 & 1.840\\ 
 6320 & 0.086  & 1.815  & 0.119 & 1.850\\ 
 7999 & 0.088  & 1.822  & 0.113 & 1.850\\ 
 9875 & 0.089  & 1.825  & 0.109 & 1.850\\
$\infty$ & 0.091(1) & 1.834(1) & 0.117(3) & 1.861(2)
 \end{tabular}
\label{table.criticalpoints}
\end{table}

\subsection{Fourfold Structure Factor} \label{sec.Sk}

\begin{figure}[htb]
 \includegraphics[width=0.8\columnwidth]{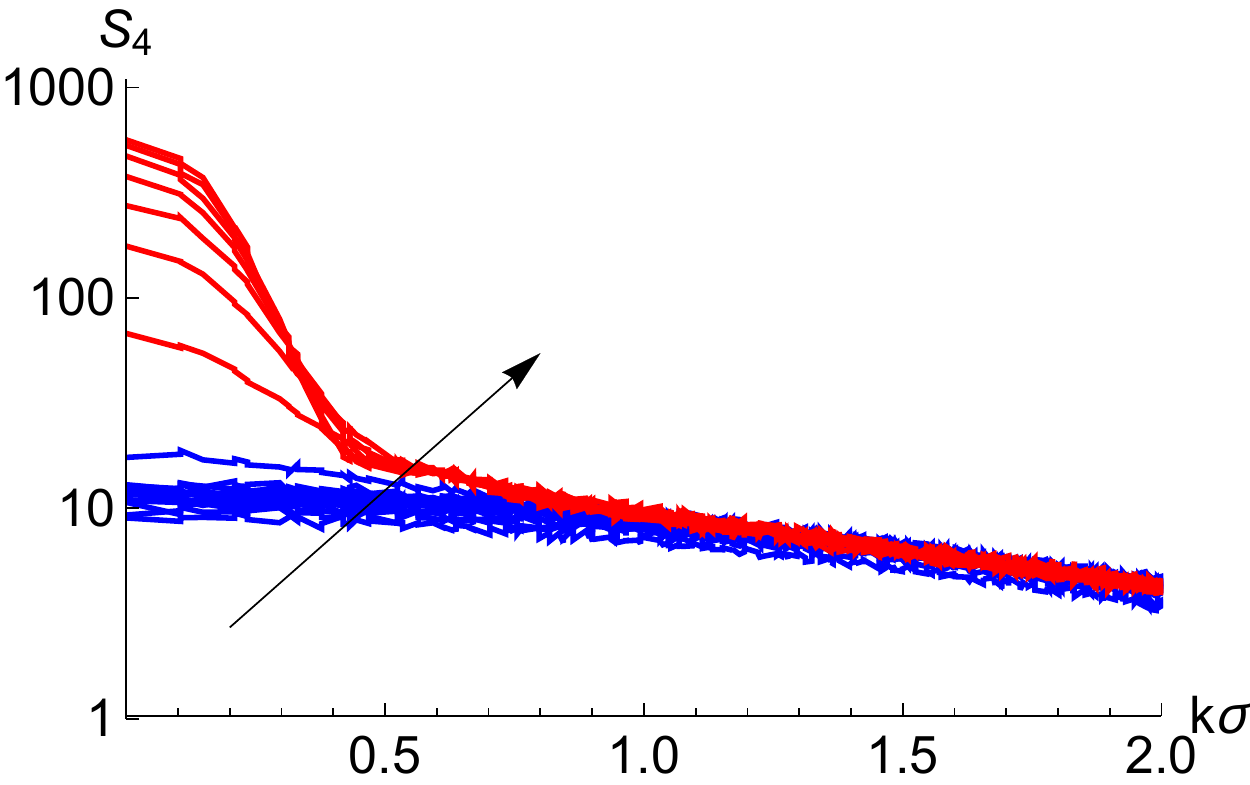}
 \caption{Fourfold structure factor $S_4(k)$ for $N=3555$ and $h=1.82\sigma$, before (blue) and after (red) the transition, for increasing values of a amplitude as indicated by the arrow. The critical amplitude is $A_c\approx 0.094 \sigma$. }
 \label{Lorentzian}
\end{figure}

\begin{figure}[htb]
 \includegraphics[width=0.49\columnwidth]{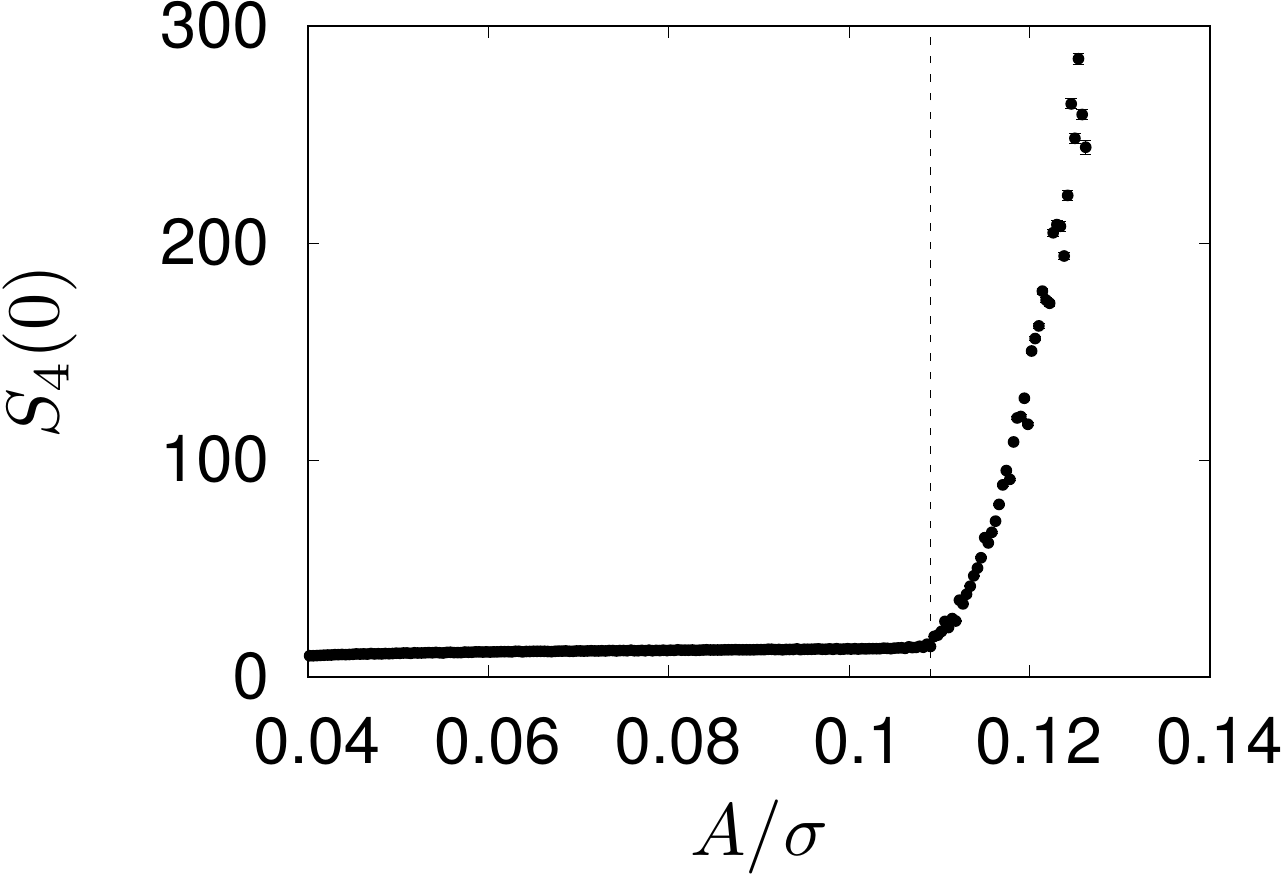}
 \includegraphics[width=0.49\columnwidth]{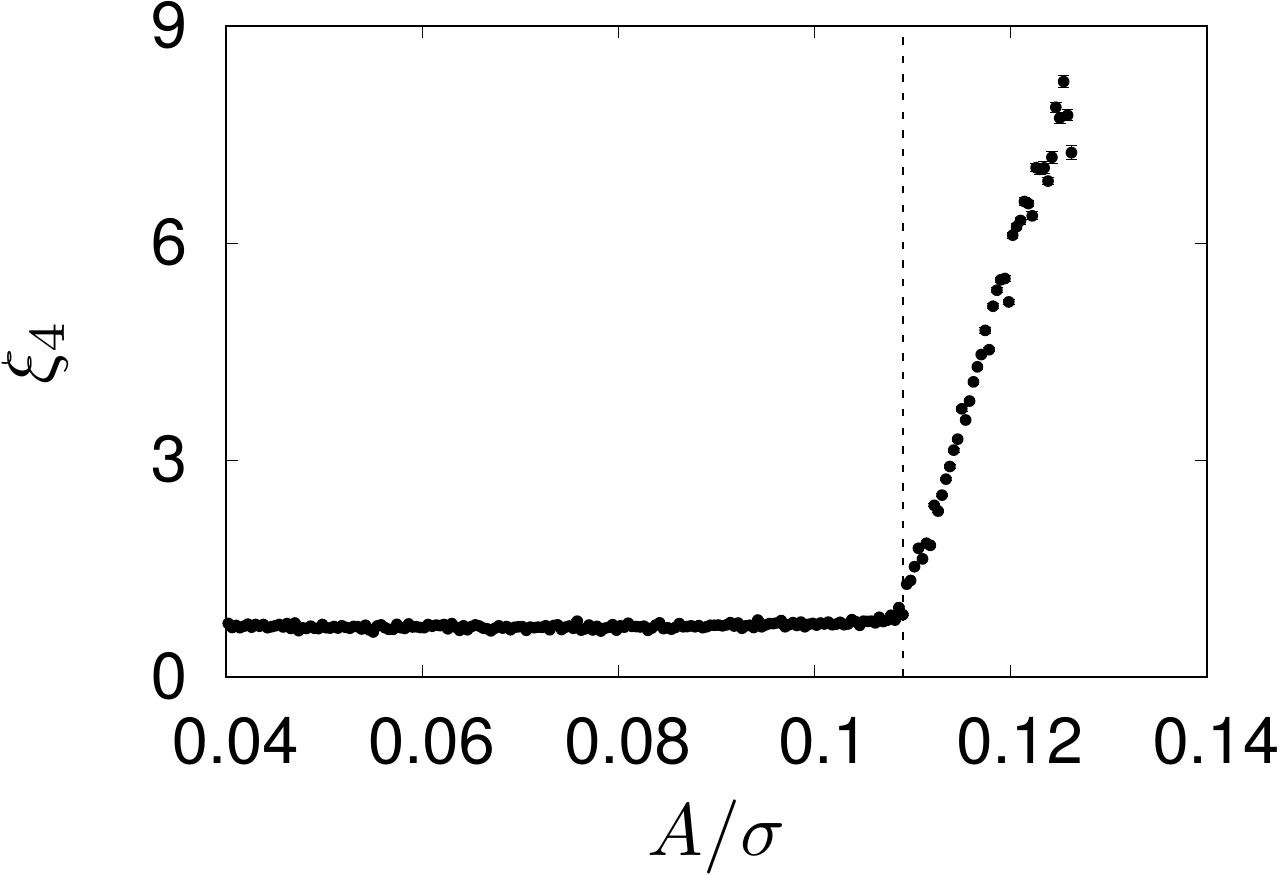}\\
 \includegraphics[width=0.49\columnwidth]{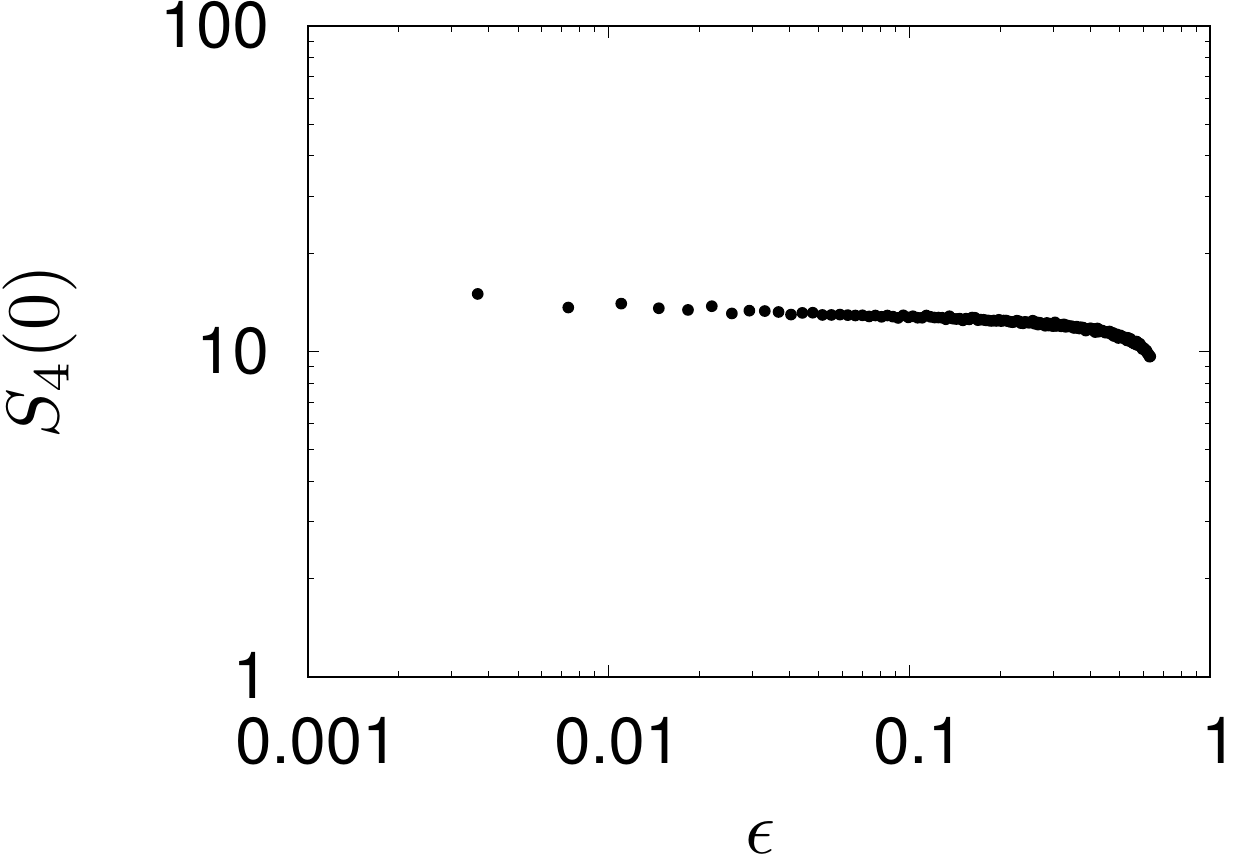}
 \includegraphics[width=0.49\columnwidth]{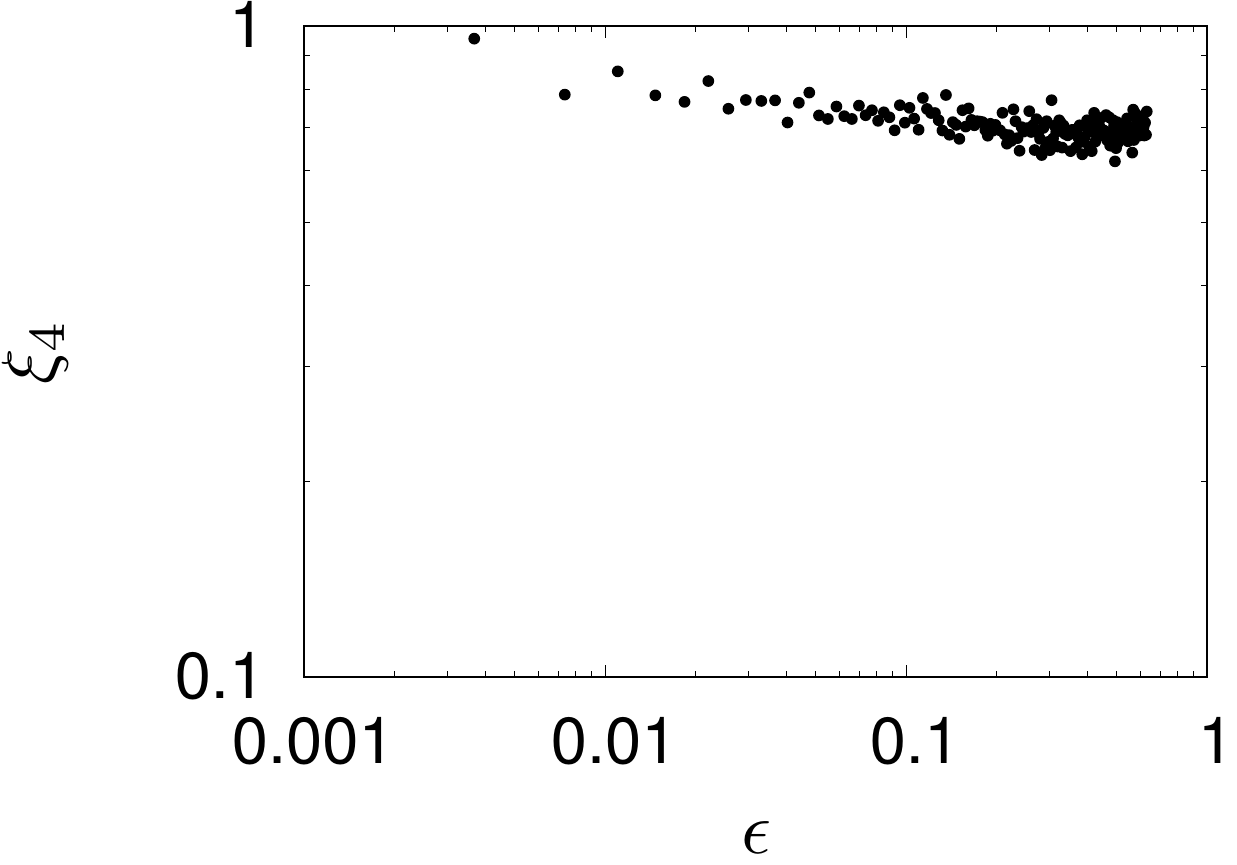}\\
 \caption{Static susceptibility $S_4(0)$ and correlation length $\xi _4$ for $N=9875$ and $h=1.85\sigma$ as a function of the amplitude $A$ (top) and the reduced amplitude $\epsilon=(A_c-A)/A_c$ in log-log scale with $A_c=0.109\sigma$ (bottom). The vertical dashed lines indicate the position of $A_c$.}
 \label{s4xi4_vsk}
 \end{figure}

We analyze the fourfold bond-orientational structure factor, $S_4(k)$, to obtain the critical properties when approaching the transition. For both kinds of transitions, an Ornstein-Zernike behavior is found in the limit of small wave number $k\sigma\ll 1$, $S_4(k)\approx S_4(0)/\left[1+(\xi_4 k)^2\right]$, as shown in Fig.~\ref{Lorentzian}. We focus our interest in the continuous case since it was found experimentally that both $S_4(0)$ and $\xi _4$ diverge, following a power law just before the transition. In order to have the largest amount of data, we analyzed the biggest system ($N=9875$) considering that $\Delta k$ scales as $1/L_{x/y}$. 
Nevertheless, neither $S_4(0)$ nor $\xi_4$ reveal any critical behavior close to the continuous transition and only present a rapid increase after the transition, due to the presence of a stable cluster, which does not correspond to critical fluctuations. Figure~\ref{s4xi4_vsk} presents both $S_4(0)$ $\xi_4$ for a box height close to the upper critical point. Similar figures are obtained for all values of $h$ between the tricritical and upper critical point.  We interpret this suppression of critical fluctuations as resulting from crossover effects of the tricritical point, which is always close to the upper critical one (see Table~\ref{table.criticalpoints}).
The same phenomena are found for the smaller systems with the exception of the smallest one ($N=1580$) for which it was not possible to fit $S_4(k)$ due to the large value of $\Delta k$.

\section{Simulations of quasi-one-dimensional systems} \label{sec.simuq1d}
\begin{figure}[htb]
 \includegraphics[width=1\columnwidth]{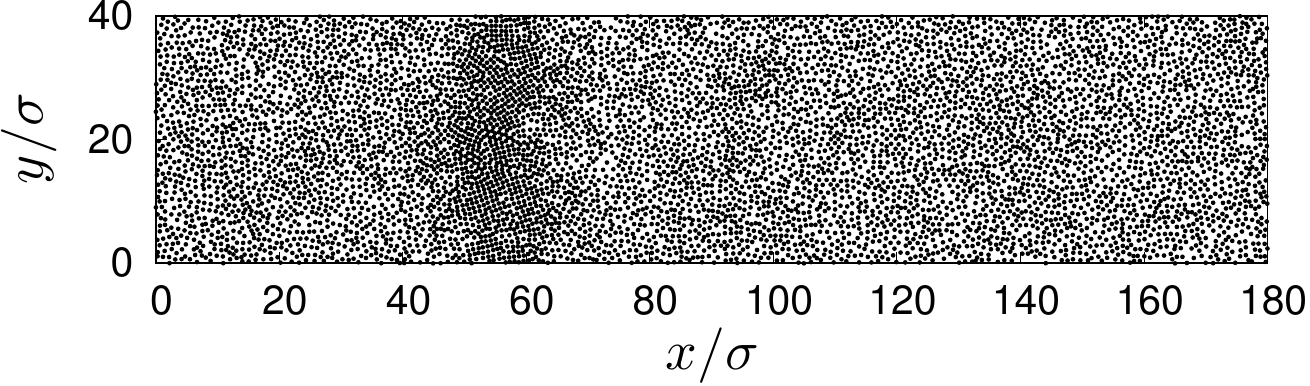}
 \caption{Typical cluster in the rectangular geometry for $h=1.8\sigma$ and $A=0.1\sigma$. The cluster consists of a rectangular strip that crosses the periodic boundary.}
 \label{rectangularCluster}
 \end{figure}

We investigate whether the lack of critical behavior in $\xi_4$ and $S_4(0)$ is due to finite size effects. To limit the computational costs of the simulation, we use a rectangular systems of dimensions $L_x=180$ and $L_y=40$, with $N=7110$, keeping  the same value for $\phi _{2D}$ as in the square systems. At the same time, this allows us to achieve smaller wavenumbers, obtaining more accurate Lorentzian fits to $S(k)$. Figure~\ref{rectangularCluster} reveals the nature of the clusters that appear in this system: they are rings in this toroidal geometry (due to the periodic boundary conditions). Rotation is practically forbidden since it would imply the rupture of the cluster, which is energetically costly.

\begin{figure}[htb]
 \includegraphics[width=0.8\columnwidth]{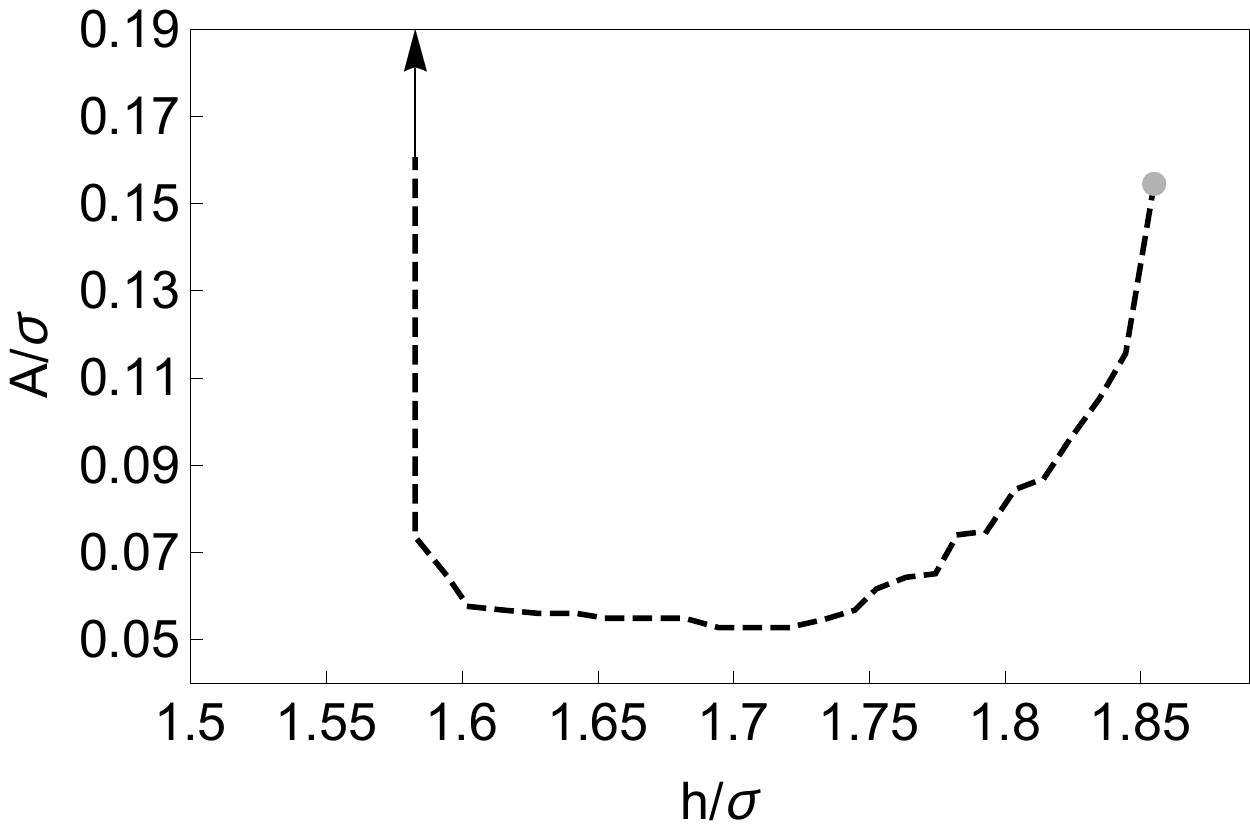}
 \caption{Phase space of the rectangular system. The dashed line indicates a discontinuous transition, ending in a critical point (gray circle). The arrow indicates that  the discontinuous transition is present up to the highest $A$, without any evidence of a lower critical point.}
 \label{ps1D}
 \end{figure}

Performing the same analysis as in the previous section, we sketch the phase space associated to the transition in Fig.~\ref{ps1D}. It is found that the continuous transition is absent, and that the discontinuous one ends up abruptly in an upper critical point. Thus, no information could be obtained regarding the critical behavior in this geometry. Other choices of the simulation parameters give consistent results.

\section{Discussion} \label{sec.discussion}

The compatibility between the discontinuous and continuous liquid to solid transitions obtained in the experiments is understood by analyzing the amplitude-height phase space using molecular dynamics simulations. A tricritical point is found in this space, where the two types of transitions merge. For heights smaller than the tricritical value, the transition is discontinuous, while for higher values the transition is continuous presenting some critical properties (with five critical laws measured in experiments, while in simulations we achieved to measure only one). The continuous transition ends in an upper critical point. For the studied parameters, the distance between tricritical and upper critical points is not large enough, resulting in important crossover effects that for large systems blur any critical behavior related to the fourfold bond-orientational structure factor $S_4(k)$ in the continuous transitions.

In this article we have also given evidence of the universality of the critical behavior associated to the fourfold bond-orientational parameter $\left<|\chi _4| \right>$. Varying the system size and the box height, and using friction coefficients different to experiments, we obtained a very robust value of the exponent $\beta =1/2$, in total agreement with the experimental results. Nevertheless, the situation is totally different regarding the fourfold structure factor, for which no critical dynamics is found near the transition point. In fact, both $S_4(0)$ and $\xi _4$ do not present divergences close to the transition.  

Rectangular systems revealed a different situation. The new topological nature of the clusters changes the type of transitions obtained in the system, eliminating the continuous one. This modification of the phase-space can be related to the effective dimensional reduction, as the rectangular geometry behaves like a quasi-one-dimensional system. We can speculate therefore, that  two is the lower dimension  in order to have a tricritical point.

\begin{acknowledgments}
This research was supported by the Fondecyt Grant No.~1140778. M.G.~acknowledges the  CONICYT PFCHA Magister Nacional Scholarship 2016--22162176.
\end{acknowledgments}

\end{document}